# Gravitational waves and primordial black holes from chirality imbalanced QCD first-order phase transition with $\mathcal{P}$ and $\mathcal{CP}$ violation

Jingdong Shao[1,*] and Mei Huang[2,†]

[1]*School of Physics, University of Chinese Academy of Sciences, Beijing 100049, China*
[2]*School of Nuclear Science and Technology, University of Chinese Academy of Sciences, Beijing 100049, China*

The chirality imbalance in QCD is spontaneously induced by a repulsive axial-vector interaction from the instanton anti-instanton pairing at high temperature above the chiral phase transition, and vanishes at low temperature. The chiral chemical potential $\mu_5$ is in the same magnitude as estimated from the sphaleron transition. Phase transition of the chirality imbalance is always a first-order one in the early universe with $\mathcal{P}$ and $\mathcal{CP}$ violation. The spectra of gravitational waves and the formation of the primordial black holes from this first-order phase transition is investigated in this work, and the effect of a strong magnetic field is also analyzed. The gravitational waves produced by chirality imbalance can be detected by LISA, Taiji and DECIGO, with the peak energy density locating in the range of $10^{-11}$ to $10^{-9}$ and the peak frequencies lying in the range of $10^{-5}$ Hz to $10^{-2}$ Hz. The spectrum with larger axial vector coupling strength and stronger magnetic field has higher peak energy density and lower peak frequency. According to this trend, the gravitational waves spectra might also be able to be detected by SKA, IPTA and EPTA. The phase transition inverse duration $\beta/H_*$ calculated from the bounce solution is in the order of $10^4$, which is much higher than typical value 10–100 from electroweak phase transitions. Based on the mechanism of postponement of the false vacuum decay, it is found that the formation of the primordial black hole is not favored because the phase transition completes in an extreme short time due to the large value of $\beta/H*$ and thus the false vacuum energy density decays sharply.

## I. INTRODUCTION

In 1916, Albert Einstein predicted that gravitational waves (GWs) can be generated by the changes of the curvature of spacetime and propagate outward at the speed of light [1,2]. 100 years later, on February 11, 2016, the LIGO and Virgo Scientific Collaboration [3] announced the first observed GWs signal named GW150914 originated from a binary black hole merge. Then the GWs signal GW170817 originated from a binary neutron star inspiral was observed by LIGO in 2017 [4].

GWs can provide an unmodified record of cosmic events, therefore the observation of GWs opens a new exciting era for astronomy and cosmology. GWs can be produced through compact binary inspirals and mergers, explosion of supernovae. Besides, it has been expected that the primordial GWs can be produced in the very early stages of the universe, which carry unique imprints on the cosmic microwave background. The spectrum of the primordial GWs can be affected by different kinds of modified gravity [5], one of which is Chern-Simons $f(R)$ gravity that ensures the chirality of the primordial GWs as discussed in Ref. [6]. GWs also can be produced from cosmological phase transitions in the early universe through electroweak (EW) and quantum chromodynamics (QCD) phase transition, and these GWs can reveal the evolution of the universe.

Current and planned GW detector projects include pulsar timing arrays, as well as ground- and space-based interferometers in the frequency range between nHz and kHz. LIGO, VIRGO [7,8] and the future Einstein Telescope (ET) [9,10] are operating in the frequency range of 1 Hz to $10^4$ Hz. 0.01–10 Hz frequency band can be covered by atom interferometers such as MAGIS ("mid-band" 30 mHz to 10 Hz) [11], AION (0.01 Hz to a few Hz) [12] and AEDGE [13] and the space-based mission the Big Bang Observer (BBO, 0.1–1 Hz) [14], DECIGO (deci-Hz, 0.1 Hz to 10 Hz) [15,16], $10^{-4} - 0.1$ Hz frequency band can be covered by space-based interferometers Laser Interferometer Space Antenna (LISA) [17], Taiji [18,19] and Tianqin [20]. The pulsar timing arrays (PTA) [21,22] including NANOGrav [23], IPTA [24] and EPTA [25], and other radio telescopes including SKA project [26,27] and the Five-Hundred-Meter Aperture Spherical Radio Telescope (FAST) project [28], can focus on the very-low-frequency GWs around 1–100 nHz through pulsar timing observation. GWs with much lower frequency have been expected to be detected through the polarization pattern of the cosmic microwave background [29] by primordial GW detectors, such as AliCPT [30], while GWs with much higher frequency covering regimes such as MHz and GHz, have been discussed in [31,32]. Most of gravitational-wave observatories and their sensitivity frequencies and energy ranges are summarized in Refs. [31,33,34].

[*]shaojingdong19@mails.ucas.ac.cn
[†]huangmei@ucas.ac.cn

A strong first-order phase transition is required to produce GWs [35–38] through the nucleation and collision of true vacuum bubbles, such as first-order inflation, GUT symmetry breaking, EW symmetry breaking as well as QCD phase transition. There is no first-order phase transition in the standard model (SM), thus first-order phase transitions beyond the standard model have been explored [39], which is also a required condition for baryongenesis (BG) [40,41]. Besides, holographic phase transitions and their corresponding GWs are discussed in Refs. [42,43].

The QCD phase transition, which is flavor and quark mass dependent [44], is rather complicated. There are two important phase transitions well defined in two extreme quark mass limits respectively, the chiral phase transition with the order parameter chiral condensation of light flavors and the deconfinement phase transition for the gluon system characterized by the order parameter of the Polyakov loop. In the chiral limit, the chiral phase transition is a second-order one in the 2-flavor case around $T_c = 175$ MeV, and a first-order one in the 3-flavor case around $T_c = 155$ MeV. The lattice calculation [45] shows that with physical quark mass, QCD phase transition at finite-temperature in the hot early universe is not a real phase transition, but a smooth crossover. When the current quark mass goes to infinity, QCD becomes pure gauge $SU(3)$ theory, and the deconfinement phase transition is of first-order at the critical temperature around $T_c = 270$ MeV. The chiral phase transition mentioned above is crucially dependent on the $U(1)_A$ axial symmetry [46]. For QCD with $N_f = 3$ massless quarks, there is an important QCD phase transition of $U(1)_A$ axial symmetry [47], which is anomalously broken due to quantum effects. This chiral anomaly is closely related to the nontrivial topological QCD theta vacuum structure [48]. The puzzle of the $\eta$ and $\eta'$ mass difference or the $U(1)_A$ problem can be resolved by instantons proposed by 't Hooft [49,50]. The topological structure of QCD vacuum is characterized by an integer-valued Chern-Simons number $N_{cs}$ [51]. Different Chern-Simons sectors are connected through instanton transitions at zero or low temperature and through sphaleron transitions at finite temperature. The change of the Chern-Simons number induces the chirality imbalance between the right-handed and left-handed quarks, which results in a violation of $\mathcal{P}$- and $\mathcal{CP}$-symmetry [52–55]. In EW theory, the sphaleron transition is associated with $B + L$ violation, with $B$ the baryon number and $L$ the lepton number [56,57].

The possibility of local parity violation at high temperatures in high-energy heavy-ion collisions as well as in the early universe was proposed as early as last century [58,59]. In recent years, QCD phase transition and phase structure under strong magnetic field have attracted much attention, because the magnetic field having a magnitude of $10^{18-20}$ G (equivalent to $eB \sim (0.1 - 1.0)$ GeV$^2$) can be generated in noncentral heavy-ion collisions [60,61] at the Relativistic Heavy Ion Collider (RHIC) or the Large Hadron Collider (LHC). It was proposed in Refs. [62–64] that under strong magnetic field, a novel phenomenon, the anomalous chiral magnetic effect (CME) can be induced by the chirality imbalance, thus an electromagnetic current can be generated along the magnetic field and the charge separation effect can be induced. The heavy-ion collisions at RHIC and LHC have made great efforts to search the CME through the observation of charge azimuthal correlations [65–67].

The essential ingredient in the CME resulting from the local $\mathcal{P}$ and $\mathcal{CP}$ violation is the chirality imbalance, which is normally introduced by an axial chemical potential $\mu_5$ [62–64]. It is shown in Refs. [68,69] that the chirality imbalance can be dynamically induced by a repulsive axial vector interaction based on the instanton–anti-instanton ($I\bar{I}$) molecule picture [70–72]. According to Refs. [70–72], the chiral phase transition can be described as a transition from instanton liquid to strongly correlated polarized instanton anti-instanton molecules. Although individual instantons and anti-instantons are strongly suppressed in the chiral symmetric phase, they keep sizable density above $T_c$ and pair up into the ordered instanton anti-instanton molecules when approaching $T_c$ from below. Therefore, the interacting instanton–anti-instanton molecules or pairing can be taken as one possible mechanism describing the non-perturbative effects of QCD in the region of $T \simeq T_c - 2T_c$ [72]. The quark interaction can be induced by the polarized instanton anti-instanton pairing above the critical temperature of the chiral phase transition $T > T_c$, and the corresponding effective Lagrangian density has been derived in Ref. [72] in the form of the four-fermion interactions similar to the Nambu-Jona-Lasinio (NJL) model. Particularly, an unconventional feature was found in Ref. [72] that the isoscalar axial-vector interaction flips sign and becomes repulsive. This repulsive axial vector interaction spontaneously induces local chirality imbalance and produces a dynamical chiral chemical potential $\mu_5$ in QCD at high temperatures above the chiral phase transition [68,69]. Also, it is found that the phase transition of the chirality imbalance is of first-order, and increases in magnetic fields help to lower the critical temperature for the appearance of the chirality imbalance. Hence the pairing of the chiral condensate is also affected by the chirality imbalance and is modified by the external magnetic fields at the temperatures around $T_c$ correspondingly. It should be emphasized that the total net topological charge should be zero in the instanton–anti-instanton molecules picture, because the number of instantons is the same as that of anti-instantons. The local chirality imbalance or local $\mathcal{P}$ and $\mathcal{CP}$ violation can be realized with one domain containing more left-handed quarks, while the other domain containing more right-handed quarks.

It is observed in Ref. [69] that the phase transition of the dynamical chirality imbalance with $\mathcal{P}$ and $\mathcal{CP}$ violation is of first-order at high temperature. A first-order phase transition is fascinating in cosmology, possibly resulting in GWs, dark matter, and baryongenesis and thus it would be very interesting to investigate the possible cosmological signatures of QCD $\mathcal{P}$ and $\mathcal{CP}$ violation. In this work we investigate the GWs spectra and the formation of the primordial black holes related to QCD $\mathcal{P}$ and $\mathcal{CP}$ violation. The primordial black holes (PBHs) have been a source of interest for nearly 50 years and can be generated through multiple approaches in different epochs [73–76], among which bubble collision during a first-order phase transition [73,77,78], especially a QCD phase transitions [79,80], is very interesting and has been widely discussed. Strong magnetic field exists in the early universe, though its origin remains a mystery. It was proposed in Ref. [81] that the primordial magnetic fields of order of magnitude of $10^{22}$ Gauss can be generated in the electroweak scale through chiral anomaly. Therefore, we will also check the effect of the magnetic field on the GWs and PBHs. The paper is organized as follows: after the Introduction, in Sec. II we introduce the model with spontaneous generation of chirality imbalance induced by a repulsive interaction in the axial vector channel, then in Sec. III, we calculate the $\alpha$ and $\beta$ parameters to obtain the GWs spectra. In Sec. IV we calculate the possibility of forming primordial black holes. At last, we give the summary and conclusion in Sec. V.

## II. AXIAL ANOMALY, CHIRALITY IMBALANCE AND FIRST-ORDER PHASE TRANSITION

The nontrivial topological configuration of QCD gauge field is characterized by the integer Chern-Simons number

$$Q_w = \frac{g^2}{32\pi^2} \int d^4 x F^a_{\mu\nu} \tilde{F}^{\mu\nu}_a \in \mathbb{Z}, \quad (1)$$

with $g$ the QCD coupling constant, $\mathrm{tr} t_a t_b = \delta_{ab}/2$ the normalized generators, $F^a_{\mu\nu}$ and $\tilde{F}^a_{\mu\nu} = \frac{1}{2}\epsilon^{\rho\sigma}_{\mu\nu} F^a_{\rho\sigma}$ the gluonic field tensor and its dual, respectively. Configurations with nonzero $Q_w$ lead to nonconservation of the axial current even in the chiral limit, which can be seen from the axial Ward-identity:

$$\partial^\mu j^5_\mu = 2\sum_f m_f \langle \bar{\psi} i \gamma_5 \psi \rangle_A - \frac{N_f g^2}{16\pi^2} F^a_{\mu\nu} \tilde{F}^{\mu\nu}_a, \quad (2)$$

with $N_f$ the number of quark flavors, $\psi$ a quark field, and $m_f$ the current quark mass, and the axial current has the form of $j^\mu_5 = \sum_f \bar{\psi}\gamma^\mu\gamma^5\psi$. In the chiral limit, we have

$$\partial_\mu j^\mu_5 = -\frac{N_f g^2}{16\pi^2} F^{a\mu\nu} \tilde{F}^a_{\mu\nu}. \quad (3)$$

The gauge field configuration with $Q_w \neq 0$ connects different topological vacua characterized by different Chern-Simons numbers, and a positive $Q_w$ convert right-handed fermions into left-handed ones

$$(N_L - N_R)_{t=\infty} = 2N_f Q_w. \quad (4)$$

Therefore, the chirality imbalance is induced by the nonzero topological charge through the axial anomaly of QCD

$$N_5 = \int d^4 x \partial_\mu j^\mu_5 = -2N_f Q_w, \quad (5)$$

with $N_5 = N_R - N_L$ denoting the number difference between right-handed and left-handed quarks, associated with the isospin singlet axial vector current. Hence, the configuration with nonzero topological charge, depending on the sign of $Q_w$, can transform left-handed quarks into right-handed or vice versa, and lead to the violation of $\mathcal{P}$- and $\mathcal{CP}$-symmetry.

The chirality imbalance density $n_5 = N_5/V$ can be represented by a chiral chemical potential $\mu_5$,

$$n_5 = \frac{\mu_5^3}{3\pi^2} + \frac{\mu_5 T^2}{3}.$$

The time evolution of $n_5$ is determined by the sphaleron transition rate $\Gamma_{ss}$,

$$\frac{\partial n_5}{\partial t} = (4N_f)^2 \frac{\Gamma_{ss}}{T} \frac{\partial F}{\partial n_5}, \quad (6)$$

where $F$ is the free energy. At zero temperature the sphaleron transition rate is tiny $\Gamma_{ss} \sim e^{-16\pi^2/g^2} \sim 10^{-160}$ and at finite temperature the sphaleron transition rate is much enhanced. Reference [82] gives $\Gamma_{ss} \sim \alpha_s^4 T^4$ for strong coupling and the chiral chemical potential can be estimated as [83]

$$\mu_5 \sim \sqrt{3}\pi \left(\frac{320 N_f^2 \Gamma_{ss}}{T^2}\right)^{\frac{1}{2}}. \quad (7)$$

When we choose $\alpha_s = 0.2$ at $T = 0.2$ GeV, we have $\mu_5 \sim 1.5$–$2.3$ GeV for $N_f = 2, 3,$ and $\mu_5 \sim 1.5$–$4$ GeV when $\alpha_s = 0.2$–$0.3$ at $T = 0.2$ GeV for $N_f = 2, 3$.

There is no full calculation on the sphaleron transition rate from high temperature to low temperature taking into account phase transitions yet, therefore in the following we will introduce a model to dynamically induce the chirality imbalance. The instanton configuration leads to the famous 't Hooft determinant among quarks with different flavors

$$\mathcal{L}_{tHooft} \sim \kappa(\det \bar{\psi}_f P_R \psi_f + \det \bar{\psi}_f P_L \psi_f), \quad (8)$$

which explicitly breaks the $U(1)_A$ symmetry due to the axial anomaly. In Eq. (8) the matrices $P_{R,L} = (1 \pm \gamma_5)/2$ are the chirality projectors. Similarly, the ordered instanton–anti-instanton pairing above the chiral critical temperature $T_c$ also results in effective quark interactions, but only in the four-fermion coupling forms. For the two-flavor case, above $T_c$ the instanton anti-instanton pairing flips the sign of the axial-vector isoscalar coupling constant from positive to negative [72], therefore one can write the interaction in the following form under magnetic field [68,69]

$$\mathcal{L} = \bar{\psi} i \gamma_\mu D^\mu \psi + G_S[(\bar{\psi}\psi)^2 + (\bar{\psi} i \gamma^5 \boldsymbol{\tau} \psi)^2] \\ - G_V(\bar{\psi}\gamma^\mu \psi)^2 - G_A(\bar{\psi}\gamma^\mu \gamma^5 \psi)^2. \quad (9)$$

Where $\tau^0$ and $\vec{\tau}$ are unit and Pauli matrices in the flavor space respectively, and $G_S$, $G_V$ and $G_A$ are the coupling constants in the scalar isoscalar, the vector isoscalar and the axial-vector isoscalar channels. $D_\mu = \partial_\mu - iq_f A_\mu$ couples quarks with electric charge $q_f$ to a magnetic field $\vec{B} = (0, 0, B)$ with vector potential $A_\mu = (0, 0, -xB, 0)$. The unconventional repulsive axial-vector interaction corresponds to a repulsive axial-vector mean field in the spacelike components but an attractive one in the timelike component.

Following Ref. [69], we only keep the scalar isoscalar and the axial-vector isoscalar channels that contribute to the phase transition significantly, and rewrite the Lagrangian density under mean field approximation

$$\mathcal{L} = -\frac{\sigma^2}{4G_S} + \frac{\mu_5^2}{4G_A} + \bar{\psi}(i\gamma_\mu D^\mu - \sigma + \mu_5 \gamma^0 \gamma^5)\psi, \quad (10)$$

where $\sigma = -2G_S\langle\bar{\psi}\psi\rangle$ is quark condensate as the order parameter of the chiral phase transition and $\mu_5 = -2G_A\langle\bar{\psi}\gamma^0\gamma^5\psi\rangle$ is the dynamical chiral chemical potential describing the chirality imbalance. In the case of $\mu_5 = 0$ and $eB = 0$, the model comes back to the regular NJL model [84], and the chiral phase transition will be a crossover at high temperature for two flavor light quarks [85].

Then the thermodynamic potential density $\Omega$ can be derived and takes the form of

$$\Omega = \frac{\sigma^2}{4G_S} - \frac{\mu_5^2}{4G_A} \\ - N_c \sum_{u,d} \frac{|q_f B|}{2\pi} \sum_{s,k} \alpha_{sk} \int_{-\infty}^{+\infty} \frac{dp_z}{2\pi} f_\Lambda^2(p) \omega_{sk}(p) \\ - 2N_c T \sum_{u,d} \frac{|q_f B|}{2\pi} \sum_{s,k} \alpha_{sk} \int_{-\infty}^{+\infty} \frac{dp_z}{2\pi} \ln(1 + e^{-\beta\omega_{sk}(p)}). \quad (11)$$

Here $p^2 = p_z^2 + 2|q_f B|k$ with $k$ a natural number marking the Landau levels, and the smooth regularization form factor [86]

$$f_\Lambda(p) = \sqrt{\frac{\Lambda^{10}}{\Lambda^{10} + p^{10}}}, \quad (12)$$

we take $\Lambda = 626.76$ MeV which satisfies $\Lambda^2 G_S = 2.02$, and use $r_A = \frac{G_A}{G_S}$ to characterize the magnitude of $G_A$ below.

The eigenvalues of the Dirac operator are

$$\omega_{sk} = \sqrt{\sigma^2 + (p + s\mu_5 \text{sgn}(p_z))^2} \quad (13)$$

with $s = \pm 1$, and the spin degeneracy factor is

$$\alpha_{sk} = 1 - \delta_{k,0} + \delta_{k,0}\delta_{s,\text{sgn}(q_f B)}. \quad (14)$$

From Eq. (11), the chiral condensate $\sigma$ and dynamical chiral chemical potential $\mu_5$ can be determined self-consistently by solving the gap equations

$$\frac{\partial \Omega}{\partial \sigma} = \frac{\partial \Omega}{\partial \mu_5} = 0. \quad (15)$$

The numerical results of chiral condensate $\sigma$ and dynamical chiral chemical potential $\mu_5$ as a function of the temperature at different $eB$ with varying $r_A$ and at different $r_A$ with varying $eB$ are shown in Figs. 1 and 2, respectively. For every case shown, the thermodynamical potential density always has two or more local minima. For simplicity, we only consider two local minima of each case, one of which indicates non-negative quark condensate $\sigma$ and zero dynamical chiral chemical potential $\mu_5$, while the other one is selected from those with positive dynamical chiral chemical potential $\mu_5$ and zero quark condensate $\sigma$ such that it has the lowest potential density $\Omega$.

From Figs. 1 and 2, it is observed that the phase transition of the chirality imbalance is always of first-order. The chirality imbalance is induced by the instanton anti-instanton pairing at high temperature, and vanishes at low temperature, which is consistent with the chirality imbalance induced by the exchange of topological Chern-Simons number through instanton transition at low temperature and

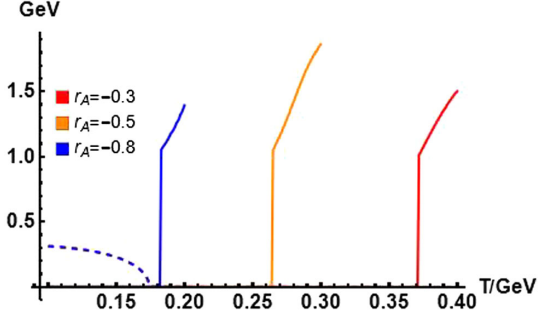

(a) $\mu_5$ (lines) and $\sigma$ (dashed lines) at $eB = 0$.

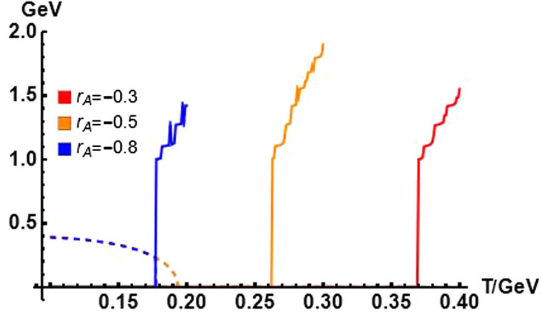

(b) $\mu_5$ (lines) and $\sigma$ (dashed lines) at $eB = 0.3 GeV^2$.

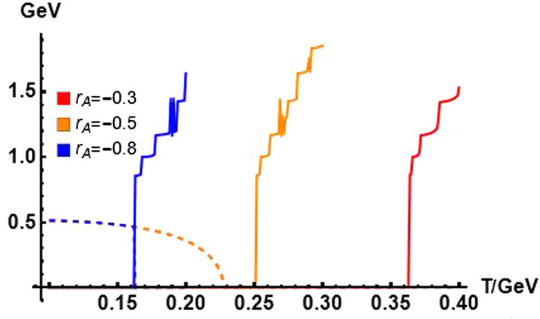

(c) $\mu_5$ (lines) and $\sigma$ (dashed lines) at $eB = 0.5 GeV^2$.

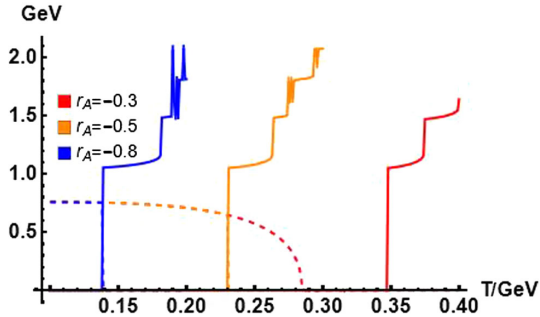

(d) $\mu_5$ (lines) and $\sigma$ (dashed lines) at $eB = 0.8 GeV^2$.

FIG. 1. Quark condensate $\sigma$ (dashed lines) and dynamical chiral chemical potential $\mu_5$ (lines) as a function of T at $eB = 0, 0.3, 0.5$ and $0.8$ GeV$^2$ for different values of $r_A$. $\sigma$ and $\mu_5$ in unit of GeV.

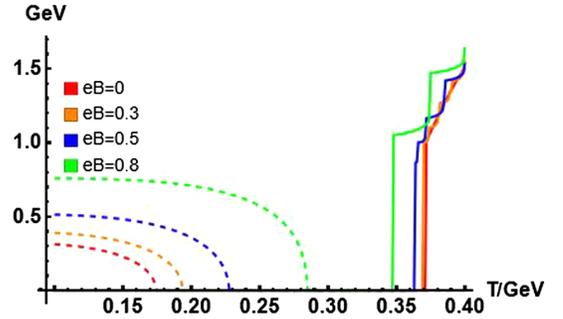

(a) $\mu_5$ (lines) and $\sigma$ (dashed lines) at $r_A = -0.3$.

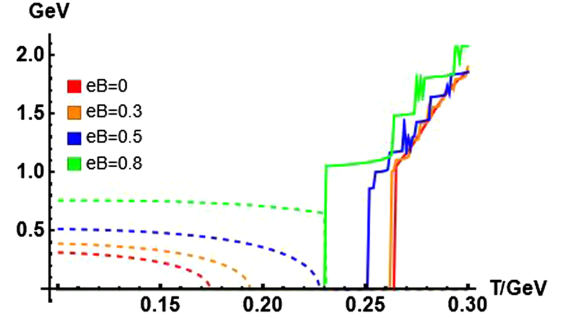

(b) $\mu_5$ (lines) and $\sigma$ (dashed lines) at $r_A = -0.5$.

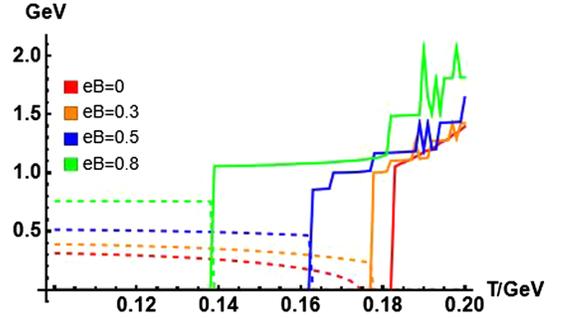

(c) $\mu_5$ (lines) and $\sigma$ (dashed lines) at $r_A = -0.8$.

FIG. 2. Quark condensate $\sigma$ (dashed lines) and dynamical chiral chemical potential $\mu_5$ (lines) as a function of T at $r_A = -0.3$, $-0.5, -0.85$ for different values of $eB$. $\sigma$ and $\mu_5$ in unit of GeV, and $eB$ in unit of GeV$^2$.

sphaleron transition at high temperature. At low temperature, the instanton transition rate is exponentially suppressed, while the sphaleron transition process is enhanced at high temperature because there is sufficient energy to pass over the barrier which separates states of different topological charge.

It is noticed that from our numerical results the induced chiral chemical potential $\mu_5$ is around 1–2 GeV above $T_c^{\mu_5}$, which is in the same order estimated from the sphaleron transition as given in Eq. (7). This indicates that our model is reasonable and can mimic chirality imbalance induced by sphaleron transition.

Figure 1 shows that the critical temperature of the chirality imbalance $T_c^{\mu_5}$ sensitively drops with stronger axial vector coupling and is not sensitive to the magnetic field. The chiral condensate and corresponding critical temperature $T_c^\sigma$ are independent of axial vector coupling strength when $T_c^{\mu_5} > T_c^\sigma$. While the magnetic field catalyzes the chiral condensate [87–89] and increases the critical temperature of chiral phase transition $T_c^\sigma$. The chiral phase transition is of second-order with small magnitude of magnetic field or/and weak axial vector coupling, and of first-order at strong magnetic field or with strong axial vector coupling. Therefore, one can observe two separated phase transitions with $T_c^{\mu_5} > T_c^\sigma$ with small magnitude of magnetic field or/and with weak axial vector coupling, when the magnitude of the magnetic field or/and the strength of the axial vector coupling increases, these two phase transitions merge into a first-order one with $T_c^{\mu_5} = T_c^\sigma$.

From Fig. 2, we can see how the magnetic field affects the phase transitions more clearly. A magnetic field $eB$ lowers $T_c^{\mu_5}$ but slightly when the axial vector coupling strength is small and raises $T_c^\sigma$ instead. Namely, $T_c^{\mu_5}$ shows an inverse magnetic catalysis effect while both chiral condensate and chiral phase transition show a magnetic catalysis effect. When the axial vector coupling strength grows, we find that the magnetic field has more influences on $T_c^{\mu_5}$ for larger absolute value of $r_A$, and another interesting result is that when the critical temperature of chiral phase transition $T_c^\sigma$ meets the one of the phase transition of chirality imbalance $T_c^{\mu_5}$, the chiral phase transition becomes first-order from a second-order one, and if the magnetic field continue to grow, the catalysis effect of magnetic field on chiral phase transition is blunted or even inverse [90–92]. This is the main result in Ref. [69] to explain the inverse magnetic catalysis, while in this work, we are interested in the first-order phase transition of the chirality imbalance. In the following, we will investigate the GWs spectra and the generation of the PBHs from this phase transition.

## III. GRAVITATIONAL WAVES

In the previous section we find that the phase transition of the chirality imbalance is always a first order one at high temperature. Once the phase transition starts, part of the universe tunnels to the true vacuum, forming bubbles with lower vacuum energy density, then the latent heat released is converted into the energy of the bubble walls. These bubbles expand and collide and pass kinetic energy to surrounding media, generating GWs from the scalar field, the sound waves and the magnetohydrodynamic (MHD) turbulence.

Growth of bubbles is dependent of the phase transition model and the bubble wall velocity can be obtained from numerical simulations [93] or from holographic perspective [94,95]. To calculate the spectra of GWs, normally the velocity of the bubble walls is taken as close to the speed of light $v_w = 0.99$. From Ref. [95], one can read the bubble wall velocity from the pressure difference. In our model, the pressure difference inside and outside the bubbles is $\Delta p = |\Omega_{\rm in} - \Omega_{\rm out}| \sim 0.005$ GeV$^4$ as shown in Fig. 3 and thus we can read the velocity of the bubble wall is about $v_w = 0.3$ from Ref. [95].

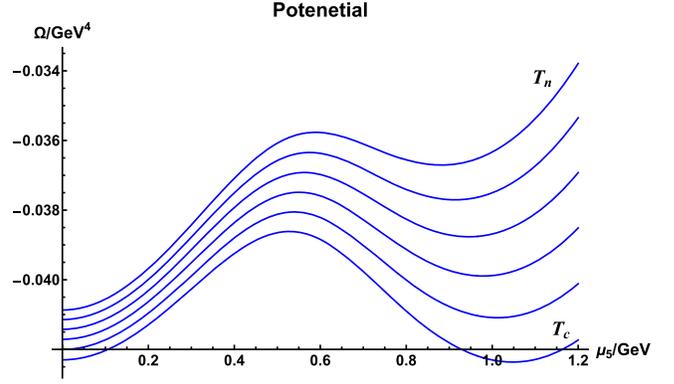

FIG. 3. $\Omega$ with $r_A = -0.5$, $eB = 0$ and $\sigma = 0$ at different temperature as a function of $\mu_5$. Temperature goes from $T_c = 0.2649$ GeV to $T_n = 0.2561$ GeV with an interval of 0.00176 GeV.

The bubble nucleation rate per Hubble volume per time is [96–98]

$$\Gamma(t) = Ae^{S_4(t)}, \quad (16)$$

where $S_4$ is Euclidean action of an $O4$-symmetric solution and reduces to $\frac{S_3}{T}$ at finite temperature $T$. And the coefficient $A$ is [99]

$$A(T) = T^4 \left(\frac{S_3}{2\pi T}\right)^{\frac{3}{2}}. \quad (17)$$

Bubbles of true vacuum start to occur when the universe cools down to the nucleation temperature $T_n$, at which we have [97,98]

$$\Gamma(t) H_*^4 \sim 1, \quad (18)$$

where $H_*$ is the corresponding Hubble parameter. Approximately $T_n$ is also the temperature of the thermal bath with weak reheating, thus we have

$$\frac{\beta}{H_*} = T_n \frac{{\rm d}(\frac{S_3}{T})}{{\rm d}T}\bigg|_{T_n}, \quad (19)$$

where parameter $\beta$ measures the transition rate.

For our model, $g_* = 12$ is the number of relativistic degrees of freedom, and hence the energy density of the radiation bath is $\rho_r = \frac{\pi^2 g_* T^4}{30}$. With the energy density difference between the false vacuum and the true vacuum $\rho_{\rm vac}$, another parameter that measures the transition strength can be written as [97,99]

$$\alpha = \frac{1}{\rho_r}\left(\rho_{\text{vac}} - \frac{T}{4}\frac{\partial \rho_{\text{vac}}}{\partial T}\bigg|_{T_p}\right)$$

$$\approx \frac{1}{\rho_r}\left(\rho_{\text{vac}} - \frac{T}{4}\frac{\partial \rho_{\text{vac}}}{\partial T}\bigg|_{T_n}\right), \quad (20)$$

where $T_p$ is the percolation temperature. As shown below, $\frac{\beta}{H_*}$ is so large that the false vacuum decays rapidly, hence the temperature is nearly constant during the phase transition, then we have $T_p \approx T_n$.

In addition, the Friedmann equation is

$$H_* = \sqrt{\frac{\rho_r + \rho_{\text{vac}}}{3m_p^2}} \quad (21)$$

with the reduced Planck Mass $m_p = 2.435 \times 10^{18}$ GeV.

Meanwhile, parameters $\kappa_v$ and $\kappa_{tb}$ are respectively the fraction of the false vacuum energy converted into the kinetic energy of the plasma and the MHD turbulence which can be analytically fitted [100–102]. Utilizing numerical fits in Ref. [101], for the wall velocity well below the sound speed $v_s$, i.e. $v_w \ll v_s$, we have

$$\kappa_A = v_w^{6/5} \frac{6.9\alpha}{1.36 - 0.037\sqrt{\alpha} + \alpha},$$

and for $v_w \sim v_s$, we have

$$\kappa_B = \frac{\alpha^{2/5}}{0.017 + (0.997 + \alpha)^{2/5}},$$

then in our subsonic case,

$$\kappa_v(v_w < v_s) = \frac{v_s^{11/5}\kappa_A\kappa_B}{(v_s^{11/5} - v_w^{11/5})\kappa_B + v_w v_s^{6/5}\kappa_A}. \quad (22)$$

Simulations give $\frac{\kappa_{tb}}{\kappa_v} \sim 0.05$–$0.1$ [100,103] and here we take $\kappa_{tb} = 0.05\kappa_v$.

In terms of the parameters above, the numerical results of GWs from sound waves and MHD turbulence are respectively [99,100]

$$h^2\Omega_{sw}(f) = 2.65 \times 10^{-6}\left(\frac{H_*}{\beta}\right)\left(\frac{\kappa_v\alpha}{1+\alpha}\right)^2\left(\frac{100}{g_*}\right)^{\frac{1}{3}}v_w S_{sw}(f) \quad (23)$$

and

$$h^2\Omega_{tb}(f) = 3.35 \times 10^{-4}\left(\frac{H_*}{\beta}\right)\left(\frac{\kappa_{tb}\alpha}{1+\alpha}\right)^2\left(\frac{100}{g_*}\right)^{\frac{1}{3}}v_w S_{tb}(f), \quad (24)$$

where

$$S_{sw}(f) = \left(\frac{f}{f_{sw}}\right)^3\left(\frac{7}{4+3(\frac{f}{f_{sw}})^2}\right)^{\frac{7}{2}}, \quad (25)$$

$$S_{tb}(f) = \left(\frac{f}{f_{tb}}\right)^3\left(1+\frac{f}{f_{tb}}\right)^{-\frac{11}{3}}\left(1+\frac{8\pi f}{h_*}\right)^{-1}. \quad (26)$$

The peak frequencies of the two sources are

$$f_{sw} = 1.9 \times 10^{-5}\frac{1}{v_w}\frac{\beta}{H_*}\frac{T_n}{100\text{ GeV}}\left(\frac{g_*}{100}\right)^{\frac{1}{6}}\text{ Hz} \quad (27)$$

and

$$f_{tb} = 2.7 \times 10^{-5}\frac{1}{v_w}\frac{\beta}{H_*}\frac{T_n}{100\text{ GeV}}\left(\frac{g_*}{100}\right)^{\frac{1}{6}}\text{ Hz.} \quad (28)$$

$$h_* = 1.65 \times 10^{-6}\frac{T_n}{100\text{ GeV}}\left(\frac{g_*}{100}\right)^{\frac{1}{6}}\text{ Hz} \quad (29)$$

is the Hubble rate.

Similarly, the other part of GWs from the scalar field is [100]

$$h^2\Omega_{\text{env}}(f) = 1.67 \times 10^{-5}\left(\frac{H_*}{\beta}\right)^2$$

$$\times \left(\frac{\kappa_\phi\alpha}{1+\alpha}\right)^2\left(\frac{100}{g_*}\right)^{\frac{1}{3}}\left(\frac{0.11v_w^3}{0.42+v_w^2}\right)S_{\text{env}}(f),$$

where

$$S_{\text{env}}(f) = \frac{3.8(f/f_{\text{env}})^{2.8}}{1+2.8(f/f_{\text{env}})^{3.8}}$$

and

$$f_{\text{env}} = 1.65 \times 10^{-5}\text{ Hz}\frac{0.62}{1.8-0.1v_w+v_w^2}\frac{T_n}{100\text{ GeV}}\left(\frac{g_*}{100}\right)^{\frac{1}{6}}.$$

Here $\kappa_\phi$ denotes the fraction of vacuum energy converted into the gradient energy of the scalar field. However, the energy in the scalar field is negligibly small for relativistic bubbles [100], only the GWs from the former two sources contribute to the total energy density spectrum, i.e.

$$h^2\Omega = h^2\Omega_{sw} + h^2\Omega_{tb}. \quad (30)$$

Table I lists nucleation temperature $T_n$ and key parameters $\alpha$ and $\frac{\beta}{H_*}$ corresponding to different values of $r_A$ and $eB$. Larger $eB$ or/and larger magnitude of $r_A$ brings lower nucleation temperature $T_n$ and thus lower peak frequencies without peaks raised according to Eqs. (27) and (28). Larger magnitude of $r_A$ also brings smaller $\frac{\beta}{H_*}$ and larger $\alpha$ while values of $eB$ have slight influence on these two parameters, and further smaller values of $\frac{\beta}{H_*}$ raise peaks according to Eqs. (23) and (24) and lower peak frequencies according to Eqs. (27) and (28), simultaneously. Additionally, $\alpha$ has no significant influence on spectra.

TABLE I. Nucleation temperature $T_n$, parameters $\alpha$ and $\beta$ corresponding to different values of $r_A$ and $eB$.

| $B/\text{GeV}^2$ | $r_A$ | $T_n/\text{GeV}$ | $\alpha$ | $\beta/H_*$ |
|---|---|---|---|---|
| 0 | −0.3 | 0.3648 | 0.7343 | 27582 |
| 0 | −0.5 | 0.2561 | 1.741 | 16274 |
| 0 | −0.8 | 0.1679 | 4.850 | 6105.7 |
| 0.3 | −0.3 | 0.3634 | 0.7727 | 30478 |
| 0.3 | −0.5 | 0.2535 | 1.790 | 14660 |
| 0.3 | −0.8 | 0.1635 | 5.375 | 12028 |
| 0.5 | −0.3 | 0.3517 | 0.7384 | 22859 |
| 0.5 | −0.5 | 0.2393 | 1.745 | 11136 |
| 0.5 | −0.8 | 0.1389 | 8.364 | 2579.0 |
| 0.8 | −0.3 | 0.3402 | 1.166 | 25235 |
| 0.8 | −0.5 | 0.2126 | 2.633 | 11171 |
| 0.8 | −0.8 | 0.1079 | 28.67 | 2819.5 |

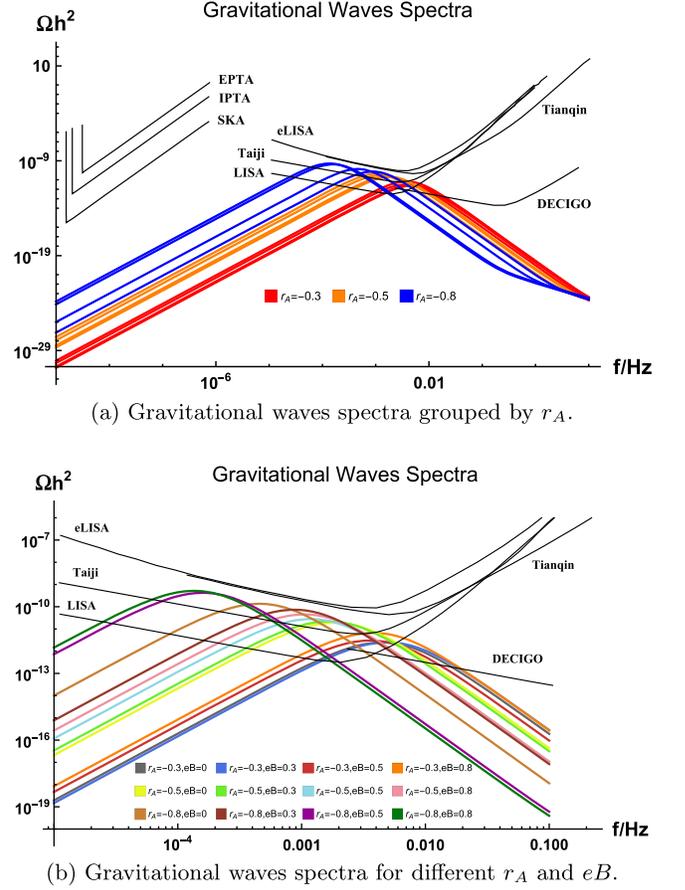

FIG. 4. Gravitational waves spectra for $r_A = -0.3, -0.5, -0.8$ with different values of $eB = 0, 0.3, 0.5, 0.8$ GeV$^2$.

It is noticed that our model gives a large value of $\frac{\beta}{H_*}$ in the order of $10^4$, and the typical value $\frac{\beta}{H_*}$ has been taken as 10–100 during electroweak phase transitions [104–106]. The parameter $\frac{\beta}{H_*}$ describes the inverse duration of the phase transition and characterizes how fast the phase transition and nucleation completed. Large value of $\frac{\beta}{H_*}$ has also been obtained in Refs. [95,107], both are from holographic QCD results. It needs to be checked in the future whether large value of $\frac{\beta}{H_*}$ is the feature of QCD phase transition and which factor determines the large value of $\frac{\beta}{H_*}$. Figure 3 shows that the shape of potential $\Omega$ changes rapidly with temperature, accounting for large values of $\frac{\beta}{H_*}$.

Figure 4 shows spectra of GWs with different values of parameters $r_A$ and $eB$. Figure 4(a) shows groups of spectra for $r_A = -0.3, -0.5, -0.8$ and detailed curves for different $eB$ can be found in Fig. 4(b). In Fig. 4(a), the peak frequencies are in the range of about $10^{-5}$ Hz to $10^{-2}$ Hz, starting around 0.001 Hz for $r_A = -0.3$ and shifting to lower values with increases in the strength of $r_A$. The energy density peaks start from around $10^{-11}$ for $r_A = -0.3$ and rise to $10^{-9}$ for $r_A = -0.8$.

Figure 4(b) shows the details for different values of $r_A$ and $eB$, where we can read the effect of the magnetic field. In general, weaker magnetic field gives higher peak frequency and lower energy density peak, with the increase in the magnetic field, the peak frequency decreases and peak energy density rises. Increasing magnitudes of both $r_A$ and $eB$ enhance the peak energy density, and the peak energy density reaches the highest in the case of $r_A = -0.8$, $eB = 0.8$.

With the parameters we consider, the GWs produced by chirality imbalance are detectable for LISA, Taiji and DECIGO [19,33], but too weak to be detectable for eLISA and Tianqin. The spectrum with larger magnitude of $r_A$ has higher peak energy density and lower peak frequency. According to this trend, if the magnitude of $r_A$ is large enough the spectrum may be able to be detected by SKA, IPTA and EPTA, while the spectrum falls to zero finally with vanishing $r_A$, meaning no first-order phase transition can be induced when the axial vector coupling constant is non-negative (see Ref. [69]).

## IV. PRIMORDIAL BLACK HOLES

Reference [108] proposes a mechanism for PBHs formation during a first-order phase transition. There probably are some areas in the universe in which the false vacuum defers decay and the postponement of decay can lead to primordial black holes. Outside such areas the phase transition starts when $\Gamma(t)H_*^4 \sim 1$ and the false vacuum energy density released is converted into other types of energy density, e.g. the radiation energy density $\rho_r$, which decays during the expansion of the universe $\rho_r \propto a^{-4}$. While inside these areas the phase transition is delayed for a certain time and the false vacuum energy density $\rho_{\text{vac}}$ stored does not decay during the expansion. Thus the energy density is larger inside than outside when the phase transition completes. If the contrast of the energy density reaches the threshold $\delta_c + 1 = 1.41$ [109], these overdense areas can collapse into PBHs. In this section, following Ref. [108], we set $8\pi G = 1$.

The bubble nucleation rate per Hubble volume per time can be expanded as [96,97]

$$\Gamma(t) = \Gamma_0 e^{\beta t}, \quad (31)$$

then the volume fraction of the remaining false vacuum is [110]

$$F(t) = \exp\left(-\frac{4\pi}{3}\int_{t_i}^{t} dt' \Gamma(t') R^3(t') r^3(t,t')\right), \quad (32)$$

where $r(t,t') = \int_{t'}^{t} dt'' \frac{v_w(t'')}{R(t'')}$ is the comoving radius of a bubble which was nucleated at time $t'$. Therefore the false vacuum energy density can be written as

$$\rho_v(t) = \rho_{\text{vac}} F(t). \quad (33)$$

Typically, we choose $\frac{\beta}{H} = 16274$ and $\alpha = 1.741$. Without significant reheating, we approximately have $\alpha = \frac{\rho_{\text{vac}}}{\rho_r(t)}$ [100].

With the energy density of the bubble walls $\rho_w$, the Friedmann equation is now

$$H = \sqrt{\frac{\rho_v + \rho_r + \rho_w}{3}}. \quad (34)$$

Then the evolution of energy density satisfies

$$\frac{d(\rho_r + \rho_w)}{dt} + 4H(\rho_r + \rho_w) = -\frac{d\rho_v}{dt}. \quad (35)$$

Figure 5 shows the evolution of the volume fraction of the false vacuum. $F(t)$ falls sharply from 1 to near 0 in a very short time, thus the phase transition progresses rapidly and we can assume the bubble nucleation rate $\Gamma_0 H_*^4 = 1$ at the initial time $t_i = 0$. The time when the first bubble occurs inside overdense areas is approximately set $t_n \Gamma_0^{\frac{1}{4}} = 0.0021$ when $F(t)$ outside is around 0.7. Since the duration of the phase transition is short in the expanding universe, $H_*$ is approximately invariant for simplicity when calculating Eqs. (31) and (32). Combining and solving Eqs. (33)–(35), we can then numerically calculate the evolution of the energy density inside and outside the overdense areas.

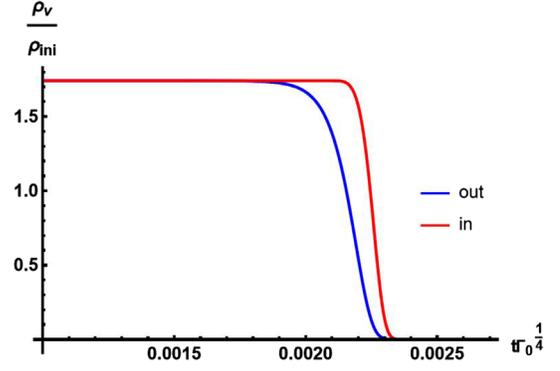

FIG. 6. The evolution of the energy density of the false vacuum inside and outside the overdense area.

Figures 6 and 7 show the inside and outside evolution of the energy density scaled by $\rho_{\text{ini}}$, the initial value of $\rho_r$. However, in our model, due to large value of $\frac{\beta}{H_*}$, the nucleation rate $\Gamma$ increases rapidly with time and the volume fraction of the false vacuum drops to near 0 rapidly, thus the false vacuum sharply decays to the true vacuum and the decay in overdense areas cannot be postponed long enough for the energy density contrast to reach the threshold. As shown in Fig. 7, the terminal energy density inside the overdense region is indeed but very slightly larger than outside, therefore given that threshold $\delta_c$ it is hardly possible to form primordial black holes through postponement of the phase transition induced by chirality imbalance without significant supercooling. If we set larger $t_n \Gamma_0^{\frac{1}{4}}$, the contrast of the energy density increases slightly but the corresponding possibility drops.

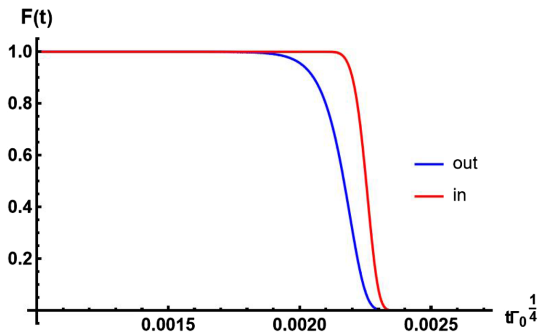

FIG. 5. The evolution of the fraction of the false vacuum.

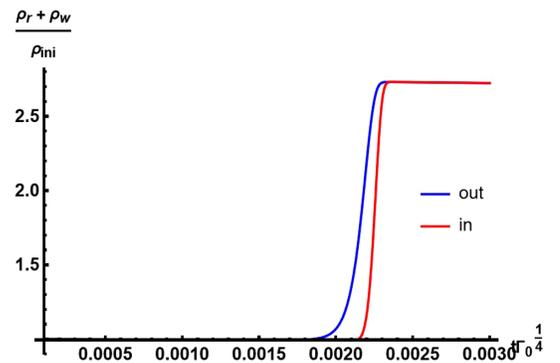

FIG. 7. The evolution of the energy density of the radiation and bubble walls inside and outside the overdense area.

## V. SUMMARY AND DISCUSSION

The nonperturbative topological structure of QCD vacuum is an important feature of pure gauge field theory, which attracts wide interest not only in the little bang in heavy-ion collisions, but also in the big bang in early universe. The topological structure of QCD vacuum characterized by integer-valued Chern-Simons number, and the change of the Chern-Simons number induces the chirality imbalance between the right-handed and left-handed number of quarks through instanton transitions at zero or low temperatures and through sphaleron transitions at high temperatures, which results in a violation of $\mathcal{P}$- and $\mathcal{CP}$-symmetry. The instanton transition rate is exponentially suppressed at low temperature, while the sphaleron transition process is enhanced at high temperature. The chirality imbalanced QCD matter with local $\mathcal{P}$ and $\mathcal{CP}$ violation related to the observation of CME in heavy-ion collisions could shed light on the mechanism of baryon asymmetry generation in the early universe [111]. In this work, we take a model with repulsive interaction in the axial vector channel derived from the instanton–anti-instanton pairing, and the chirality imbalance can be spontaneously induced at high temperature above the chiral phase transition, and vanishes at low temperature. We investigate the cosmological observations related to QCD phase transition with $\mathcal{P}$ and $\mathcal{CP}$ violation.

It is found that the phase transition of the chirality imbalance is always a first-order phase transition, and the critical temperature of the chirality imbalance $T_c^{\mu_5}$ is sensitively lowered by larger magnitude of the axial vector coupling constant but not much by stronger magnetic field. The chiral phase transition is of second-order with weak magnetic field and small magnitude of the axial vector coupling constant and in this situation the chiral condensate is catalyzed by the magnetic field with corresponding critical temperature $T_c^{\sigma}$ raised but independent of the axial vector coupling constant. It becomes first-order with strong magnetic field and large magnitude of axial vector coupling constant and in this situation, if the magnetic field continue to grow, the chiral condensate is less or inversely catalyzed with decreasing critical temperature keeping $T_c^{\sigma} = T_c^{\mu}$. Therefore, two phase transitions are separated when $T_c^{\mu_5} > T_c^{\sigma}$, when the magnitudes of the magnetic field and axial vector coupling constant are large enough, the chiral phase transition becomes first-order with $T_c^{\mu_5} = T_c^{\sigma}$ so that these two phase transitions merge into a first-order one and the magnetic catalysis effect is blunted or inverse.

Then we investigate the generation of GWs from this first-order phase transition and consider the effect of a strong magnetic field. We calculate the corresponding parameters $\alpha$ and $\beta$ for different values of parameters $r_A$ and $eB$ and obtain the GWs spectra. We present how the shape of the potential changes as the temperature decreases to explain large values of $\frac{\beta}{H_*}$ in the order of $10^4$. Large $\frac{\beta}{H_*}$ means that the phase transition completes in a short time, which is unusual compared with typical values $\frac{\beta}{H_*} \sim 10$–$100$ during electroweak phase transitions. It also remains uncertain whether large $\frac{\beta}{H_*}$ is a unique character of QCD phase transitions and which factor determines this feature. We select two figures of spectra to demonstrate changes of spectra with $r_A$ or $eB$. Weaker magnetic field gives higher peak frequency and lower peak energy density, with the increase in the magnetic field, the peak frequency decreases and peak energy density rises. Increasing magnitudes of both $r_A$ and $eB$ enhance the peak energy density, and the peak energy density reaches the highest in the case of $r_A = -0.8, eB = 0.8$. The GWs spectra produced by chirality imbalance are detectable for LISA, Taiji and DECIGO, with the peak energy density locating in the range of $10^{-11}$–$10^{-9}$, and the peak frequency lying in the range of $10^{-5}$ Hz to $10^{-2}$ Hz. The spectrum with larger magnitude of axial vector coupling constant and stronger magnetic field has higher peak energy density and lower peak frequency. According to this trend, the GWs spectra might also be able to be detected by SKA, IPTA and EPTA.

At last, we investigate the possibility of formation of the PBHs from this first-order phase transition based on the mechanism of postponement of false vacuum decay. We calculate the evolution of the energy density inside and outside those overdense areas, and find that with typical parameters the phase transition completes in a too short time because of large value of $\frac{\beta}{H_*}$ and thus the false vacuum energy density decays sharply. The postponement cannot last long while keeping feasible possibility, hence the contrast of the energy density inside and outside overdense areas can hardly reach the threshold 1.41 and it is scarcely possible to form PBHs in our model. We show that this mechanism becomes invalid during the QCD phase transition with $\frac{\beta}{H_*} \sim 10^4$.

With typical parameters above, the phase transition induces GWs that can only be detected by LISA, Taiji and DECIGO. However, if the phase transition occurs with much stronger supercooling, the situation becomes different and intriguing. Lower transition temperature means larger $\alpha$ and much smaller $\frac{\beta}{H_*}$. If strong supercooling occurs alone, e.g. $\frac{\beta}{H_*} \sim 10$, the peak of the spectrum moves to the upper left and thus the GWs can also be detected by SKA, IPTA and EPTA with peak frequencies $f \sim 10^{-7}$ Hz and peak energy density around $10^{-7}$. If a weaker supercooling happens, e.g. $\frac{\beta}{H_*} \sim 1000$, and simultaneously $T_n$ is higher when we set tiny magnitude of $r_A$, the peak then rises and the GWs are detectable for Tianqin and eLISA.

Furthermore, if supercooling is strong enough to reach $\frac{\beta}{H_*} \sim 1$, the decay of the false vacuum can be postponed longer enough (e.g. $t_n \Gamma_0^{\frac{1}{4}} \sim 1$), so that the contrast of the energy density inside and outside can probably reach the threshold and it is possible to form PBHs. However, in our model, we have $\frac{\beta}{H_*} \sim 10^4$, the nucleation rate $\Gamma$ increases rapidly with time and the volume fraction of the false vacuum drops to near 0 rapidly, thus the false vacuum sharply decays to the true vacuum and the decay in overdense areas cannot be postponed long enough for the energy density contrast to reach the threshold. If we want $\frac{\beta}{H_*}$ to drop from the order $10^4$ to the order of 1 in our model, extreme strong supercooling is required. This demands some part of the universe remains the false vacuum in spite of the nucleation rate $\Gamma(t)H_*^4 \gg 1$ and thus the corresponding possibility of that strong supercooling is extremely slight. Though postponement of the false vacuum decay has negligible possibility to form PBHs in this model, other mechanisms may work, e.g. bubble collisions [78,112], density fluctuations [113]. Meanwhile, if there exist a dustlike phase which lowers acoustic speed $v_s$ during a QCD phase transition, the threshold $\delta_c$ decreases and make it easier to form PBHs [79,114].

As we have shown in our model, the mechanism of postponement of the false vacuum decay proposed in [108] is not general to form PBHs for any first-order phase transitions as the authors claimed. In their calculations they chose two sets of typical parameters $\alpha = 6, \beta/H_* = 14.8$ and $\alpha = 0.5, \beta/H_* = 3.7$ from electroweak phase transitions. Our results show that large value of $\beta/H_*$ do not favor the formation of PBHs. If the large value of $\beta/H_*$ is a feature of QCD phase transitions with strong coupling, it is not good news for PBHs as one of the candidates of dark matter.

The physics of $\mathcal{P}$ and $\mathcal{CP}$ violation induced by sphaleron transition is fundamental, and the predicted chiral magnetic effect (CME) has attracted much attention and has been an important target in heavy ion collisions. We expect that $\mathcal{P}$ and $\mathcal{CP}$ violation induced by QCD sphaleron transition will leave some imprints in the early universe, for example, it might contribute to the baryogenesis and chiral GWs, and induce possible observables of $\mathcal{P}$ and $\mathcal{CP}$ violation in CMB [115]. We will leave these interesting topics for future studies.

## ACKNOWLEDGMENTS

We thank Ligong Bian, Yidian Chen, Huaike Guo, Mingqiu Li, Jing Liu and Qishu Yan for helpful discussions. This work is supported in part by the National Natural Science Foundation of China (NSFC) Grants No. 12235016, No. 12221005, No. 11725523 and No. 11735007, the Strategic Priority Research Program of Chinese Academy of Sciences under Grants No. XDB34030000 and No. XDPB15, the start-up funding from University of Chinese Academy of Sciences (UCAS), and the Fundamental Research Funds for the Central Universities.